\documentstyle[aps,prd,epsf]{revtex}
\begin{document}
\title{Color Singlet Strangelets}
\author{Dan M{\o}nster Jensen, Jes Madsen and Michael B. Christiansen}
\address{ Institute of Physics and Astronomy \\
University of Aarhus \\
DK-8000 {\AA}rhus C, Denmark\\[15pt]
Talk presented at: Strangeness`96 May 15-17, Budapest, Hungary}
\date{June 1996}
\maketitle
\begin{abstract}
When considering strangelets at finite temperature it is important to
obey the constraint that any observed state must be color singlet.
This constraint results in an increase in strangelet masses as
calculated at fixed entropy per baryon. We use the color singlet
partition function for an MIT bag, derived using the group
theoretical projection method, to calculate strangelet masses.
Mean shell effects are included in a liquid drop model, by using a
density of states obtained from the multiple reflection expansion.
Another important effect of the color singlet restriction, namely many
orders of magnitude suppression of thermal nucleation of quark-gluon
plasma in collisions, will also briefly be described.
\end{abstract}
 
\section{Introduction}
The (meta-) stability of strange quark matter as suggested by several
authors \cite{Bodmer71a,Chin79a,Terazawa79a,Witten84a} has spurred
theoretical as well as experimental activities aimed at answering the
question of whether strange quark matter already exists in Nature 
({\it e.g}, in the cores of neutron stars or in cosmic rays) or
can be produced in a laboratory (most likely in relativistic heavy-ion
collisions). It is not an easy question to answer with our present
theoretical tools, since it involves non-perturbative aspects of the
strong interaction. Precisely for this reason most of the theoretical
work has been done using different phenomenological models. Ultimately
the question of the stability of strange quark matter will have
to be answered experimentally/observationally, {\it e.g}, by the
detection of strangelets in heavy ion collisions.
Strangelet production in ultrarelativistic heavy-ion collisions would
also constitute an unambiguous signal of quark-gluon plasma
formation.

While not making specific predictions for the outcome of an experiment
here, we do try to address some of the points which are believed to be
important for experimental strangelet searches. These are the effects of
small system size and the fact that the produced
strangelet is not born cold, but has to survive the hot
environment
ensuing the collision of two heavy nuclei.  In addition to the added
thermal energy, which in itself results in an increase of strangelet masses,
there is a more subtle effect, {\it viz.} the fact that the strangelet
must be color singlet.

We have used the MIT bag model \cite{Chodos74a}  to describe a
strangelet as an ideal Fermi gas of $u$, $d$ and $s$ quarks (and also
antiquarks and gluons at finite temperature). We use a density of
states derived using the multiple relection expansion \cite{Balian70a}
which gives corrections due to the finite system size.
This is the liquid drop model of strangelets, similar to the liquid drop
model of nuclei.
The overall effects of the shells are well described by the liquid drop
model, whereas the detailed effects of the shells, such as shell
closures, which are likely to be important at very low baryon number are not
revealed in the liquid drop model. However,  recent results
\cite{Mustafa96a} indicate that shell effects are washed out at
elevated temperatures, making shell model calculations and liquid drop
calculations effectively  equivalent.  The main advantage of the liquid
drop model over the shell model is that many quantities can be
evaluated analytically.

Some of the results presented
here are described in Ref.~\cite{Jensen96a}, where only massless quarks were
treated.

\section{Strangelets at Finite Temperature}
\begin{figure}
   \centering
   \epsfxsize=\textwidth
   \epsfbox{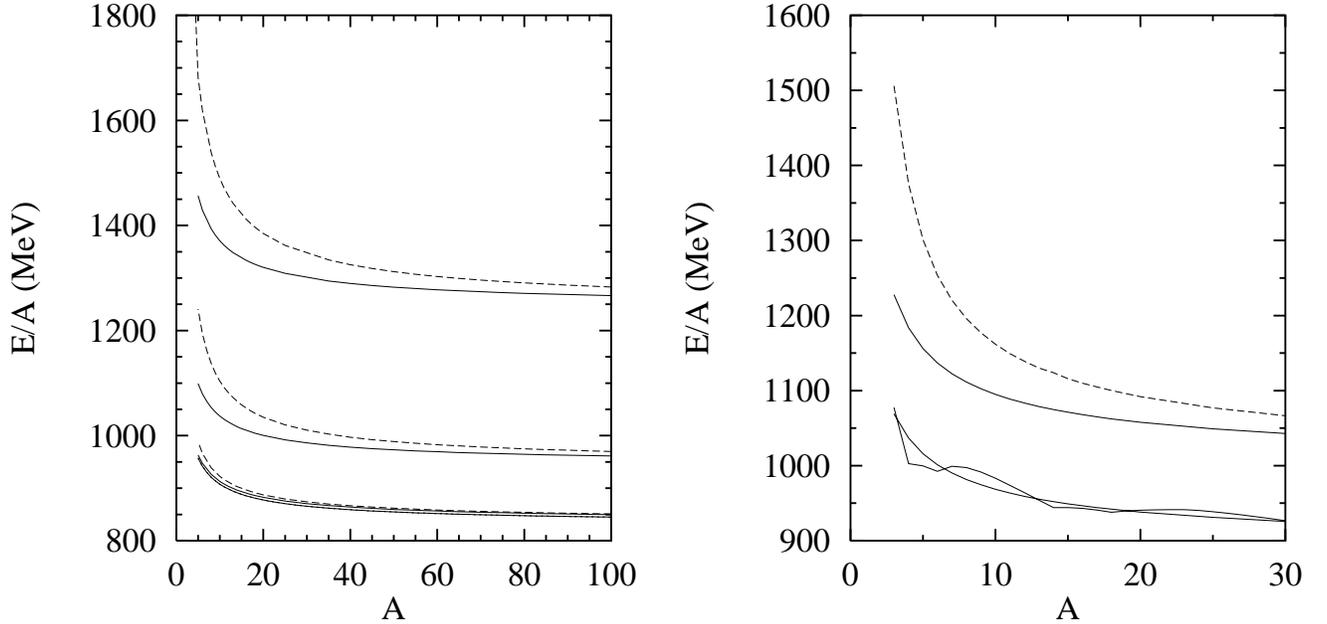}
   \caption{The energy per baryon number as a function of baryon number.
	    Full curves show results without the color singlet
	    constraint---dashed curves with it.  {\bf Left}:
	    curves for fixed entropy per baryon (from top to bottom:
	    $S/A=$ 10,5,1,0) with three massless flavors.  {\bf Right}:
	    Equivalent results for a strange quark mass of 150
	    MeV. The two sets of curves correspond to $S/A=$ 0 and 5.
	    The jagged curve is a shell model calculation. In both
	    panels $B^{1/4}=145$ MeV. \label{fig:energy} }
\end{figure}

We first consider strangelets at non-zero temperatures without
inclusion of the color-singlet constraint. In this case we can obtain
all relevant information from the grand canonical partition function $Z$
of a noninteracting gas of quarks, antiquarks, and gluons at temperature
$T=\beta^{-1}$ in an MIT bag of volume $V$
\begin{equation} \label{eq:lnZ}
	\ln Z = \sum_i \ln Z_i -\beta BV.
\end{equation}
Here $Z_i$ is the partition function for particle species $i$ and $B$ is
the bag constant, which is the energy density of the perturbative vacuum
inside the bag.
Quarks, antiquarks, and gluons contribute terms of the form
\begin{equation} \label{eq:lnZi}
	\ln Z_i = \pm g_i \int_0^\infty dk\, \rho(k)
		\ln\left\{ 1 \pm \exp\left[ -\beta\left( \epsilon(k) - \mu_i
		\right) \right] \right\},
\end{equation}
where the upper sign is for fermions, and the lower for bosons. $g_i$ is
the statistical weight due to the spin (helicity) and color degrees of
freedom, $\epsilon(k)$ is the energy of a particle with momentum $k$, and
$\mu$ is the chemical potential. We use units in which $\hbar=k_B=c=1$.
The density of states $\rho(k)$ is
a smooth function of $k$ calculated in the framework of the multiple
reflection expansion \cite{Balian70a}. It has the
general form
\begin{equation}
	\rho(k) = \frac{Vk^2}{2\pi^2} + f_S\left(\frac{m}{k}\right) Sk
		+ f_C\left(\frac{m}{k}\right) C + \cdots,
\end{equation}
where $S$ is the surface area of the bag, and $C\equiv\int_S dS (1/R_1 +
1/R_2)$ is the extrinsic curvature of the bag surface ($R_1$ and $R_2$
are the principal radii of curvature). The functions $f_S$ and $f_C$
depend on the equations of motion and on the boundary conditions, but
\emph{not} on the geometry of the confining surface, which is only assumed
to be smooth. The surface coefficient for a quark field in a cavity with
confining MIT bag boundary conditions is
\cite{Berger87a,Mardor91a}
\begin{equation}
        f_S\left(\frac{m}{k}\right) = -\frac{1}{8\pi} \left( 1
-\frac{2}{\pi}
                \tan^{-1} \frac{k}{m} \right).
\end{equation}
Note that this expression vanishes in the limit $m \to 0$, so that
massless quarks do not contribute to the surface tension. So for massless
($u$ and $d$) quarks the lowest order correction to the density of
states is the curvature term.
The curvature coefficient $f_C$ has not been calculated using the multiple
reflection expansion in the general case of massive quarks, but
for massless quarks with MIT bag boundary conditions the result is
\cite{Mardor91a}
$f_C = -1/24\pi^2$. In the limit of infinite mass the quark can be
considered non-relativistic and the field equation reduces to the wave
equation. In this case the boundary condition of the MIT bag is
the well known Dirichlet boundary condition studied by Balian and Bloch
\cite{Balian70a} who derive the result $f_C = 1/12\pi^2$. It was shown
by Madsen \cite{Madsen94a} that the expression
\begin{equation} \label{eq:curvature}
        f_C\left(\frac{m}{k}\right) = \frac{1}{12\pi^2} \left[ 1 -
        \frac{3k}{2m} \left(\frac{\pi}{2} - \tan^{-1} \frac{k}{m}\right)
       \right],
\end{equation}
which has the above special cases as limits, is in very good agreement
with shell model calculations (see also Fig.~\ref{fig:energy}).
We will therefore adopt
(\ref{eq:curvature}) as the curvature coefficient for quarks.
For gluons the relevant expressions were given by Mardor and Svetitsky
\cite{Mardor91a}
\begin{equation}
        f_S = 0 \qquad \qquad f_C = -\frac{1}{6\pi^2}.
\end{equation}
Just as massless quarks, gluons do not contribute to the surface
tension.

It is possible to calculate the thermodynamical potential $\Omega =
-T\ln Z$ without resorting to numerical evaluation in the two limits:
$T \to 0$ and $m \to 0$. Since we are mainly interested in the finite
temperature regime we will give the result for ${\cal N}_q$ massless
quark flavors, including antiquarks and gluons, and assuming a common
chemical potential $\mu_q$ of all quarks (antiquarks have
$\mu_{\bar{q}} = -\mu_q$) 
\begin{eqnarray} \label{eq:omega}
	\Omega(T,\mu_q,V,C) = &-& \left[ \left( \frac{7{\cal N}_q}{60}
	+ \frac{8}{45} \right) \pi^2T^4 + \frac{{\cal N}_q}{2} \left(
	\mu_q^2T^2 + \frac{\mu_q^4}{2\pi^2} \right) -B \right] V
	\nonumber \\
	&+& \left[ \left( \frac{{\cal N}_q}{24} + \frac{4}{9} \right)
	T^2  + \frac{{\cal N}_q}{8\pi^2}\mu_q^2 \right] C.
\end{eqnarray}
In the following we consider only spherical systems, where $V = 4\pi R^3
/3$ and $C = 8\pi R$ are given in terms of the radius $R$. When fixing
the geometry in this manner there is only one independent ``shape
variable'' which we will take to be the volume.
A strangelet is in mechanical equilibrium when $\partial \Omega/
\partial V|_{\mu_q,T} = 0$. The chemical potential and the temperature
are determined by requiring the baryon number $A$ and the entropy per
baryon $S/A$ to be fixed. So to find the equilibrium values of $T$,
$\mu_q$ and $V$ we simultaneously solve the equations
\begin{equation} \label{eq:ligevaegt}
	\left(\frac{\partial \Omega}{\partial V}\right)_{\mu_q,T} = 0,
	\quad\qquad
	\left(\frac{\partial \Omega}{\partial \mu_q}\right)_{T,V} = 3A,
	\quad\qquad
	-\frac{1}{A} \left(\frac{\partial \Omega}{\partial T}\right)_{\mu_q,V}
	= \frac{S}{A}
\end{equation}
The energy (mass) $E$ of the strangelet can then be calculated by use of the
identity
\begin{equation}
	E = \Omega + TS + 3\mu_qA.
\end{equation}
As long as only massless quarks are considered this is equivalent to the
expression $E = 4BV$ \emph{in equilibrium}. This is because massless
particles satisfy the relativistic
equation of state $E = 3PV$, where $P$ is the pressure. In equilibrium
the dynamic pressure of the particles is exactly balanced by the bag
constant $B = P$, so that adding the vacuum energy $BV$ gives $E = 3BV +
BV = 4BV$. Including a massive strange quark this no longer holds.
Plots of the energy per baryon as a function of baryon number obtained
by solving Eqs.~(\ref{eq:ligevaegt}) are shown as full lines in
Fig.~\ref{fig:energy}.

\section{Color Singlet Strangelets}
The partition function given by Eqs.~(\ref{eq:lnZ}) and (\ref{eq:lnZi})
is in fact the grand canonical partition function for a colorless
assembly of quarks, antiquarks, and gluons in an MIT bag. In order to
force a non singlet color configuration it is
necessary to introduce chemical potentials related to the color degree
of freedom of the quarks and gluons. Two chemical potentials,
corresponding to the two conserved charges (Cartan operators) of $SU(3)$,
are needed.
Now we need to put in a sum over the different color states of quarks
and gluons in the partition function, thus replacing Eq.~(\ref{eq:lnZi})
with
\begin{equation} \label{eq:lnZicolor}
	\ln Z_i = \pm g'_i \sum_c \int_0^\infty dk\, \rho(k)
		\ln\left\{ 1 \pm e^{ -\beta\left[ \epsilon(k) - \mu_i
		-\gamma_1C_1(c) - \gamma_2C_2(c) \right]  } \right\}.
\end{equation}
The statistical weight $g'_i$ no longer includes the color degree of
freedom, which is now explicitly summed over. $C_1$ and $C_2$ are the
eigenvalues of the Cartan operators, which for quarks are in the
fundamental representation and for gluons in the adjoint representation.
Using the group theoretical projection method developed by Redlich and
Turko \cite{Redlich80a,Redlich80b,Turko81a}, it is possible to construct
a canonical partition function  (with respect to color) from the grand
canonical $Z(\gamma_1,\gamma_2)$ constructed using the prescription
just outlined.
The resulting color-singlet canonical partition function is
\begin{equation} \label{eq:Zsinglet}
	Z_{\rm C.S.} = \int_{-\pi}^{\pi} d\theta_1 d\theta_2\,
	M(\theta_1,\theta_2) Z(iT\theta_1,iT\theta_2),
\end{equation}
where $M(\theta_1,\theta_2)d\theta_1d\theta_2$ is the Haar measure
of $SU(3)$. Note that
the chemical potentials $\gamma_i = iT\theta_i$ are now purely
imaginary. The baryon number is still treated in a grand canonical
ensemble, so $Z_{\rm C.S.}$ is a mixed canonical/grand canonical
ensemble.

In the thermodynamic limit ($V\to\infty$) the two ensembles are
equivalent, {\it i.e.}, they give identical mean values---differing only
with respect to fluctuations. For a finite system this is no longer the
case, and since color neutrality is exact at all times the canonical
prescription is the proper one.

For massless quarks it is possible to evaluate $Z_{\rm C.S.}$
analytically in a saddle-point approximation, valid at high $T$ and/or
$\mu_q$. This was done by Elze and
Greiner \cite{Elze86a}, who for ${\cal N}_q$ massless quark flavors
obtained the result
\begin{equation} \label{eq:ZCS}
	Z_{\rm C.S.}(T,V,\mu_q) = \Pi(T,V,\mu_q) Z(T,V,\mu_q),
\end{equation}
where $\Pi$ is given by
\begin{equation} \label{eq:PI}
        (2\pi\sqrt{3}\, \Pi)^{-1/4} =
        VT^3\left\{2 + {\cal{N}}_q\left[ \frac{1}{3}
	+\left( \frac{\mu_q}{\pi T} \right) ^2 \right] \right\}
        + CT \frac{12 - {\cal{N}}_q}{12\pi^2}.
\end{equation}
Note that $\Pi$ vanishes in the limit $V\to\infty$. The limit $T\to 0$
cannot be taken in the above expression for $\Pi$, since the
saddle-point approximation breaks down for low $T$. But for $T\to 0$ the
effect of exact color singletness also disappears since at $T=0$ it is
always possible to construct a color singlet state for $3A$ quarks
occupying the lowest energy levels available.
For massive quarks we evaluate the partition function numerically.

The effect of the color singlet constraint on the energy per baryon is
seen in Fig.~\ref{fig:energy}, where dashed lines include the color
singlet constraint. There is a sizeable effect regardless of whether one
includes a massive strange quark or not. The main effect of the strange
quark mass is to shift curves towards higher $E/A$. Curves for zero
strange quark mass are calculated using the saddle-point approximation,
which has been shown to be a good approximation in this case (see
Ref.~\cite{Jensen96a}).

\section{Color Singlet Suppressed Nucleation}
As another example of the effect of color singletness we consider the
nucleation of quark matter droplets from hadronic matter using standard
homogenous nucleation theory. The low baryon number density region
of the phase diagram
is where the transition to a quark-gluon plasma would happen in
RHIC and LHC experiments. In the region of the phase diagram
characterized by high baryon number density and moderate temperature
this could be relevant for the nucleation of quark matter droplets in
AGS experiments and in
the core of neutron stars. In the case of neutron stars the time scales are
probably sufficently large for homogenous nucleation theory to be
applicable, whereas this is questionable in the case of
quark-gluon plasma formation in heavy ion collisions.
An estimate for the nucleation rate is
\begin{equation} \label{eq:rate}
	{\cal R} \approx T^4 \exp\left( -\Delta F/T \right),
\end{equation}
where $\Delta F$ is the height of the free energy barrier that has to be
surmounted by the nucleated bubble. For the prefactor we have chosen the
dimensional estimate $T^4$. One could consider improvements to this
estimate but since the relative suppression due to
color singlet effects are dominated by the exponential, this is unlikely
to change the results significantly.

\begin{figure}
   \leavevmode\epsfxsize=0.48\textwidth\epsfbox{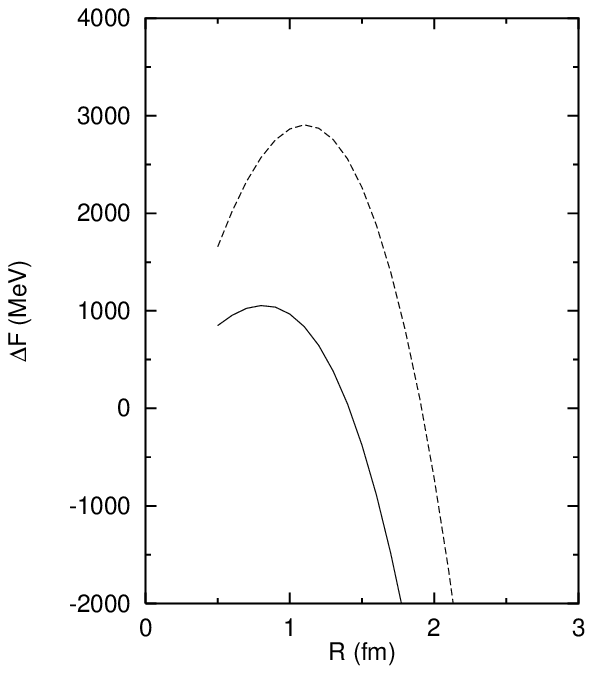}\hfill
   \epsfxsize=0.48\textwidth\epsfbox{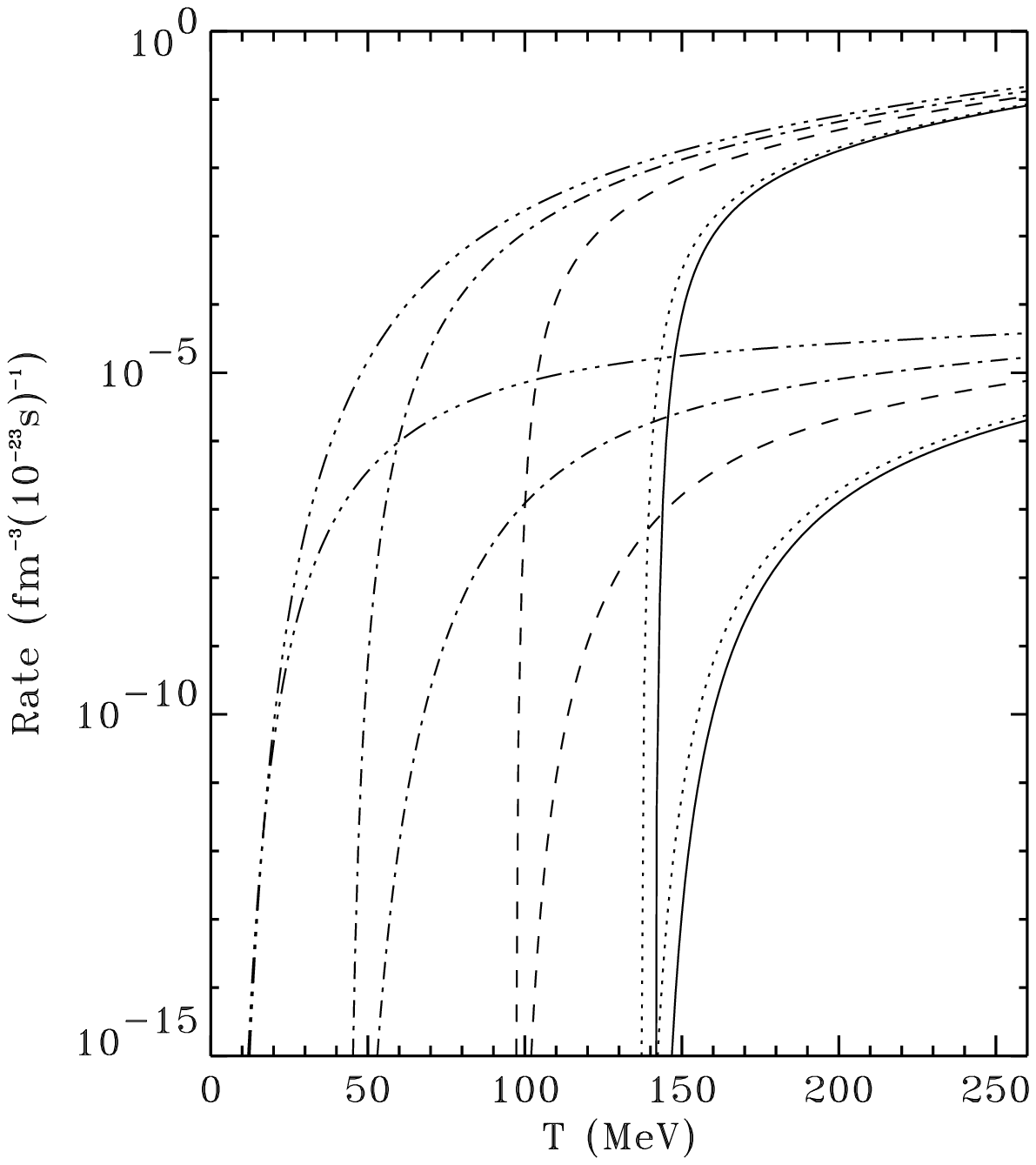}
   \caption{\label{fig:barrier}
            {\bf Left}: Free energy barrier for $T=170$ MeV and $\mu_q=0$.
	    The upper curve is with the singlet constraint, the
	    lower is not. {\bf Right}: Nucleation rate
	    for $\mu_q= 0$ (solid), 100 
	    (dotted), 300 (dashed), 400 (dot-dashed), and 500 MeV
	    (dash-triple dot). Lower curves with singlet
	    constraint, upper without. In both panels $B^{1/4}=200$ MeV.
	    }
\end{figure}

The free energy barrier is given by the pressure difference
between the hadronic phase and the plasma phase (times volume)
plus additional contributions from surface tension and the curvature
term.
\begin{equation}
	\Delta F =  (P_{\rm h} - P_{\rm q})V + \sigma S + \gamma C.
\end{equation}
Note that any contribution from the hadronic phase, {\it i.e.},
either a non-zero pressure or a surface tension, will increase the
barrier height. We will therefore ignore the hadron gas contribution,
since it will only augment the effect due to color
singletness (for a more elaborate discussion of this point, see
Ref.~\cite{Madsen96a}). With this simplification the free energy barrier
is simply $ \Delta F = \Omega$, where $\Omega$ is given by
Eq.~(\ref{eq:omega}) with ${\cal N}_q=2$. The color singlet constraint
is taken into account by using $\Omega = -T\ln Z_{\rm C.S.}$ as given by
Eqs.~(\ref{eq:ZCS}) and (\ref{eq:PI}). The barrier is shown for a
particular choice of parameters in
Fig.~\ref{fig:barrier}.

Inserting the value of $\Delta F$ for a critical bubble in the
expression (\ref{eq:rate}) for a range in temperature and chemical
potential gives a plot like the one shown in the right panel of
Fig.~\ref{fig:barrier}, where an extra constraint, namely requiring the
droplet to have zero momentum \cite{Madsen96a}, is also included.
The effect of the color singlet constraint, which is the dominant of
the two constraints, is seen to be quite dramatic. It causes a
suppression of the plasma formation rate by four to five orders of
magnitude.

\section{Conclusions}

We have shown that the effect of exact color singletness plays an
important role in strangelets---increasing masses as calculated at fixed
entropy per baryon---as well as in the phase transition between hadronic
and quark matter, where it causes a suppression of quark gluon plasma
nucleation.

\vskip 10pt
\noindent \large {\bf Acknowledgement}\normalsize
\vskip 10pt
\noindent
DMJ gratefully acknowledges the hospitality of Brookhaven National Lab.\
where part of the work described here was done.

This work was supported in part by the Theoretical Astrophysics Center,
under the Danish National Research Foundation, and the U.S.\ Department
of Energy under contract DE-AC02-76CH00016.

\vfill\eject 
\end{document}